\documentclass{ametsocV6.1}

\usepackage{hyperref}
\usepackage{acro}
\usepackage{placeins}
\usepackage{booktabs}  
\usepackage{multirow}
\usepackage{subcaption}
\usepackage[colorinlistoftodos]{todonotes} 

\graphicspath{ {./Figures/} }

\DeclareAcronym{iwc}{ 
    short = {IWC}, 
    long  = {ice water content},
}
\DeclareAcronym{iwp}{ 
    short = {IWP}, 
    long  = {ice water path},
}
\DeclareAcronym{nice}{
    short = {N$_{\textrm{ice}}$},
    long  = {ice crystal number concentration},
}
\DeclareAcronym{ttl}{
    short = {TTL},
    long  = {tropical tropopause layer},
}
\DeclareAcronym{inp}{
    short = {INPs},
    long = {ice nucleating particles},
}

\DeclareAcronym{mcs}{
    short = {MCS},
    long  = {mesoscale convective system},
}
\DeclareAcronym{itcz}{
    short = {ITCZ},
    long  = {intertropical convergence zone},
}
\DeclareAcronym{ir}{
    short = {IR},
    long  = {infrared},
}
\DeclareAcronym{arm}{
    short = {ARM},
    long  = {Atmospheric Radiation Measurement},
}
\DeclareAcronym{ml}{
    short = {ML},
    long  = {machine learning},
}
\DeclareAcronym{cnn}{
    short = {CNN},
    long = {convolutional neural network},
}
\DeclareAcronym{mlp}{
    short = {MLP},
    long = {multilayer perceptron}
}
\DeclareAcronym{gan}{
    short = {GAN},
    long = {generative adversarial network},
}
\DeclareAcronym{vae}{
    short = {VAE},
    long = {variational auto encoder},
}
\DeclareAcronym{ddpm}{
    short = {DDPM},
    long = {denoising diffusion probabilistic models},
}
\DeclareAcronym{relu}{
    short = {ReLU},
    long = {rectified linear unit},
}
\DeclareAcronym{gelu}{
    short = {GELU},
    long = {Gaussian error linear unit},
}

\newif\ifarxiv
\arxivtrue

\title{IceCloudNet: 3D reconstruction of cloud ice from Meteosat SEVIRI}

\authors{Kai Jeggle,\aff{a}\correspondingauthor{Kai Jeggle, kai.jeggle@env.ethz.ch}
Mikolaj Czerkawski,\aff{b}
Federico Serva,\aff{b,c}
Bertrand Le Saux,\aff{b}
David Neubauer,\aff{a}
Ulrike Lohmann,\aff{a}\correspondingauthor{Ulrike Lohmann, ulrike.lohmann@env.ethz.ch}}

\affiliation{
    \aff{a}{Institute for Atmospheric and Climate Science, ETH Zurich, Zurich, Switzerland} \\
    \aff{b}{$\Phi$-lab, European Space Agency (ESA), Frascati, Italy} \\
    \aff{c}{Consiglio Nazionale delle Ricerche - Istituto di Scienze Marine (CNR-ISMAR), Rome, Italy}  
}

\ifarxiv\hypersetup{
pdftitle={IceCloudNet: 3D reconstruction of cloud ice from Meteosat SEVIRI},
pdfsubject={physics.ao-ph, cs.AI},
pdfauthor={Kai Jeggle, Mikolaj Czerkawski, Federico Serva, Bertrand Le Saux, David Neubauer, Ulrike Lohmann},
pdfkeywords={Clouds, Cirrus, Mixed-Phase, 3D reconstruction, Cloud Structure, Machine Learning, AI},
}

\abstract{IceCloudNet is a novel method based on machine learning able to predict high-quality vertically resolved cloud ice water contents (IWC) and ice crystal number concentrations (N$_{\textrm{ice}}$). The predictions come at the spatio-temporal coverage and resolution of geostationary satellite observations (SEVIRI) and the vertical resolution of active satellite retrievals (DARDAR). IceCloudNet consists of a ConvNeXt-based U-Net and a 3D PatchGAN discriminator model and is trained by predicting DARDAR profiles from co-located SEVIRI images. Despite the sparse availability of DARDAR data due to its narrow overpass, IceCloudNet is able to predict cloud occurrence, spatial structure, and microphysical properties with high precision. The model has been applied to ten years of SEVIRI data, producing a dataset of vertically resolved IWC and N$_{\textrm{ice}}$ of clouds containing ice with a 3 km$\times$3 km$\times$240 m$\times$15 minute resolution in a spatial domain of 30°W to 30°E and 30°S to 30°N. The produced dataset increases the availability of vertical cloud profiles, for the period when DARDAR is available, by more than six orders of magnitude and moreover, IceCloudNet is able to produce vertical cloud profiles beyond the lifetime of the recently ended satellite missions underlying DARDAR.}

\begin{document}

\maketitle


\statement
    Clouds containing ice remain a source of great uncertainty in climate models and future climate projections. IceCloudNet overcomes the limitations of existing satellite observations and fuses the strengths of high spatio-temporal resolution of geostationary satellite data with the high vertical resolution of active satellite retrievals through machine learning. With this work we are providing the research community with a fully temporal and spatial resolved 4D dataset of cloud ice properties enabling novel research ranging from cloud formation and development to the validation of high-resolution weather and climate model simulations.

%

\section{Introduction}

Clouds containing ice in the tropics cover over 20 \% of Earth's surface at any moment \citep{heymsfield_cirrus_2017}. These clouds can be categorised into two temperature-based regimes: cirrus clouds occur at temperatures below -38°C containing only ice crystals and mixed-phase clouds that occur between 0°C and -38°C and may simultaneously contain supercooled cloud droplets and ice crystals \citep{lohmann_introduction_2016}. Mixed-phase clouds are mainly formed by mid-level or deep convection in the tropics. In the case of deep convection, cirrus clouds can form at the top of convective towers with the typical anvil shape. Cirrus clouds can also form directly from the vapor phase at high altitudes and temperatures below -38°C (in situ cirrus), e.g. caused by gravity waves \citep{gasparini_opinion_2023}. All clouds exert a radiative effect by modulating the incoming solar and outgoing terrestrial radiation \citep{liou_influence_1986}. The radiative effect is driven by the microphysical properties of the cloud and its macrophysical structure. Cirrus clouds are typically thinner and have on average a warming effect \citep{heymsfield_cirrus_2017}, while mixed-phase clouds are thicker and usually exert a cooling effect \citep{lecuyer_reassessing_2019}. Due to their radiative effect, their impact on global circulation and regional precipitation, tropical clouds containing ice are of a large climatological significance. \Acp{mcs} for instance, which consist of multiple convective cores and corresponding anvil clouds, are responsible for the majority of precipitation in the tropics \citep{feng_global_2021}. Despite the climatological significance of cirrus and mixed-phase clouds in the tropics, there are gaps in our understanding of cloud formation and evolution, which lead to large uncertainties in climate projections \cite{forster_earths_2021}. The cloud response of anvils in a warming climate, for instance, represents the largest uncertainty of all cloud feedbacks \citep{sherwood_assessment_2020,forster_earths_2021} and has recently gained momentum in the research community \citep{mckim_weak_2024, sokol_greater_2024}. Cloud ice properties are projected to change in a warming climate and may amplify or dampen global warming \citep{lohmann_importance_2018}. For a comprehensive review of tropical cirrus clouds we refer to \cite{gasparini_opinion_2023}. A main challenge of studying clouds are processes that act on a wide range of spatio-temporal scales. The evolution of anvil cirrus, for instance, depends on mesoscale and cloud-scale circulations driven by radiative and latent heatings, which in turn depend on microphysical properties. As a result, anvil cirrus can exert a cooling or warming effect, depending on their life cycle. To improve process understanding, which is key to reduce uncertainties, it is necessary to exert a perspective on formation and evolution when studying clouds \citep{heymsfield_cirrus_2017, gasparini_opinion_2023} in contrast to studying individual temporal snapshots of clouds, which is usually done in observational studies. To study cloud formation and development holistically, a three-dimensional spatial view is required to follow the structural cloud evolution, in addition to the evolutionary perspective.

Key microphysical properties to characterize cirrus and mixed-phase clouds are \ac{iwc} and \ac{nice}, which denote the ice mass and ice crystal number concentration per unit volume of air respectively. Given \ac{iwc} and \ac{nice} also the effective particle radius of ice crystals can be calculated. \ac{iwc} and effective particle radius are mainly determining the radiative properties of a cloud. \ac{nice} on the other hand can provide insight into aerosol-cirrus interactions \citep{gryspeerdt_ice_2018, mitchell_calipso_2018} and ice nucleation pathways \citep{gryspeerdt_automated_2018}.

Satellite-based remote sensing retrievals are mainly used for studying clouds, next to modelling and in-situ studies. We focus in this study on two of the main types of satellite retrievals used for studying clouds. On the one hand, multiple studies \citep[e.g.][]{kramer_microphysics_2016, sourdeval_satellite-based_2020} have used polar-orbiting active satellite instruments like CALIPSO's lidar \citep{winker_caliop_2009} and CloudSat's radar \citep{stephens_cloudsat_2002} to analyze microphysical properties of ice clouds. In this work, we will use the DARDAR (raDAR/liDAR) \citep{delanoe_variational_2008,cazenave_evolution_2019} dataset which combines the CALIPSO and CloudSat instruments to retrieve cloud profiles of \ac{iwc} and in its extension, DARDAR-Nice, also \ac{nice} \citep{sourdeval_ice_2018}. On the other hand, passive geostationary satellite instruments such as the Spinning Enhanced Visible and Infrared Imager (SEVIRI) onboard the Meteosat satellites \citep{aminou_msgs_2002} enable a perspective on  the temporal evolution of clouds. Such instruments retrieve a top-down 2D view of Earth's surface every 15 minutes by measuring the reflected solar radiation and emitted longwave radiation and hence enable a long-term view on clouds with a high spatio-temporal resolution and coverage.  

\Ac{ml} has evolved has a powerful tool to extract information and knowledge from earth observation data \citep{tuia_artificial_2024}. Previous studies have employed \ac{ml} models of various complexity to predict actively sensed retrievals from geostationary satellite data. \cite{kox_retrieval_2014} have used a simple fully connected neural network and \cite{amell_ice_2022} a \ac{cnn} to predict \ac{iwp}, which is the vertically integrated \ac{iwc}, from SEVIRI. \cite{amell_chalmers_2023} extended the approach to also predict \ac{iwc} from a single \ac{ir} channel globally by combining multiple geostationary satellites with a 1 km vertical resolution. \cite{leinonen_reconstruction_2019} applied a \ac{gan} based approach to reconstruct vertical cloud profiles from polar-orbiting satellite instruments (MODIS). 


\begin{figure}[ht]
  \centerline{\includegraphics[width=39pc]{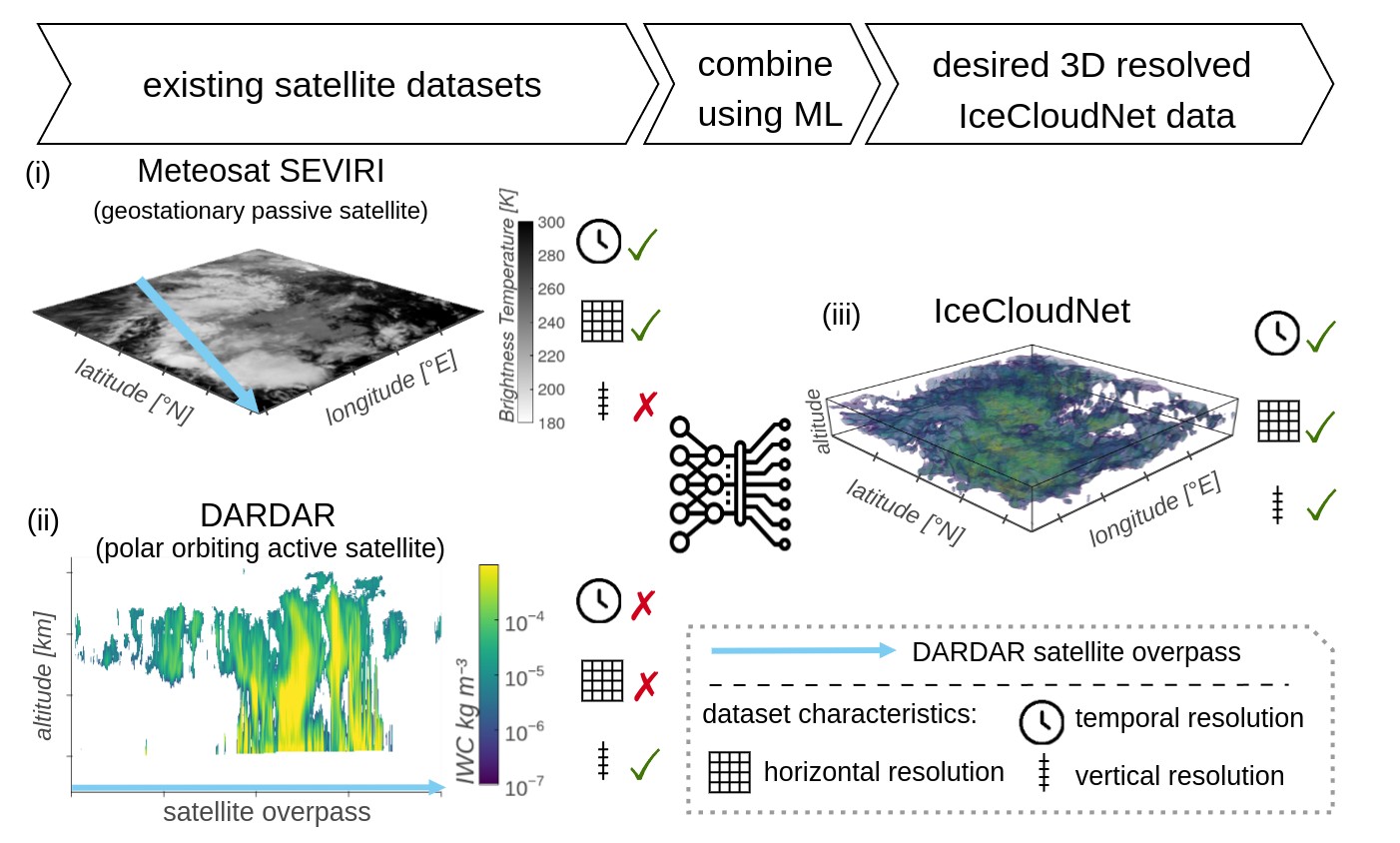}}
  \caption{Conceptual visualization of the idea behind IceCloudNet. Panel (i) shows a sample of Meteosat SEVIRI and panel (ii) the vertical \ac{iwc} profile along the co-located DARDAR satellite overpass. The strengths and weaknesses of both datasets are indicated by the icons next to the panels. Through the use of \ac{ml} a novel synthetic dataset is created (iii), which combines the spatio-temporal resolution and coverage of SEVIRI with the vertical resolution of DARDAR. The shown data is from 2006-08-22 13:12:40 in a spatial domain ranging from 3°W to 7°E, 9°N to 19°N.}
  \label{fig:concept}
\end{figure}

The objective of this work is to combine the vertically resolved view on cloud ice properties of DARDAR with the high spatio-temporal resolution and coverage of SEVIRI as visualized in Fig.~\ref{fig:concept}. To this end, we train a \ac{cnn} to predict \ac{iwc} and \ac{nice} for cirrus and mixed-phase clouds from SEVIRI data, which is supervised by co-located DARDAR retrievals.



We note that this work is an extension to the conference workshop paper \citep{jeggle_icecloudnet_2023}, where the authors presented a first version of IceCloudNet capable of predicting \ac{iwp} from SEVIRI data.

\section{Data}\label{sec:data}
\subsection{Training and test datasets}\label{sec:data_train}

The dataset utilized for training and validating IceCloudNet comprises six years (2007-2012) of multi-spectral images from Meteosat Second Generation (MSG) SEVIRI and vertically-resolved DARDAR \ac{iwc} and \ac{nice} swaths, where the former is used as input for IceCloudNet and the latter as prediction targets.

All available SEVIRI channels in the visible and \ac{ir} spectrum are used as input for IceCloudNet. The three visible channels are dependent on the scattering of solar radiation and are hence only available for daytime predictions. The eight \ac{ir} channels are sensitive to thermal emission of the observed atmosphere and surface and are thus available during day and night. The fact that different channels are sensitive to different penetration depths (visible) and emission levels (\ac{ir}) motivates the use of all available channels as input to IceCloudNet.

\begin{figure}[h]
 \centerline{\includegraphics[width=27pc]{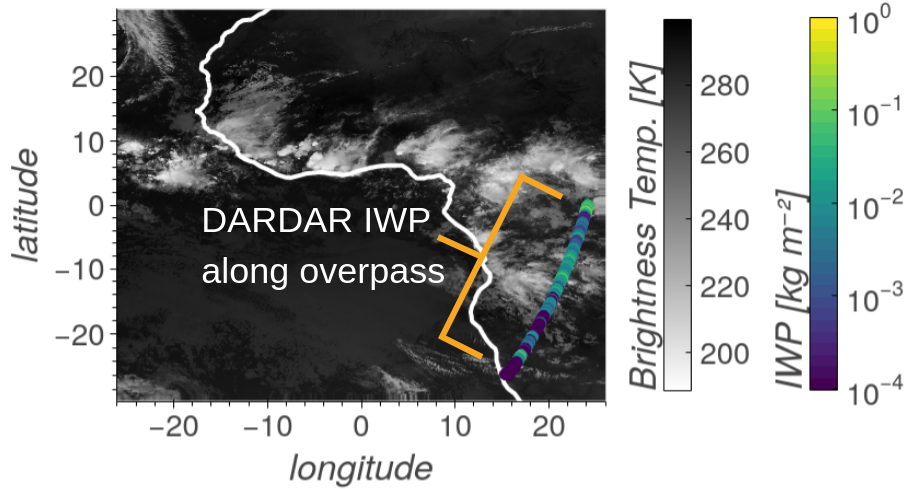}}
  \caption{10 \textmu m channel for single SEVIRI observation from 2007-10-24 00:12:43 with co-located DARDAR IWP. The DARDAR swath is magnified for better visibility.}\label{fig:data_viz}
\end{figure}

For the \ac{iwc} retrievals, the latest version (v3) of DARDAR is employed \citep{cazenave_evolution_2019}, and for \ac{nice}, its extension DARDAR-Nice \citep{sourdeval_ice_2018} in its second version is utilized. Throughout this study, we use \ac{nice} with effective radii $>$ 5 \textmu m. Both \ac{iwc} and \ac{nice} have been thoroughly validated and, despite the intrinsic uncertainties of satellite retrievals, show good agreement with in-situ measurements \citep{kramer_microphysics_2020, cazenave_evolution_2019, sourdeval_ice_2018}. DARDAR has a resolution along the satellite overpass of 1.2 km at the surface with an observation every 0.16 seconds. For every DARDAR overpass, we identify and co-locate the matching SEVIRI image by calculating the minimum difference between DARDAR and SEVIRI observation times. Given the 15-minute temporal resolution of SEVIRI and the actual observation time being 12.5 minutes, the maximum difference between co-located SEVIRI and DARDAR observations is 6.25 minutes. While this timescale is tolerable for most cloud processes, it introduces a bias in our co-located data. Once co-located, the DARDAR data is resampled and averaged to the 3 km x 3 km native resolution of SEVIRI. Due to the temporal sampling every 15 minutes and the long revisiting times of DARDAR, any SEVIRI image contains at most one DARDAR overpass, resulting in a minor fraction of a given SEVIRI image that can be co-located with DARDAR data. Note that many SEVIRI images do not contain any DARDAR overpass. Fig.~\ref{fig:data_viz} visualizes an exemplary SEVIRI image with its matching DARDAR swath. To reduce the data size and required computational resources, we resample and average the DARDAR data to a vertical resolution of 240 m from its original 60 m resolution. A vertical resolution of 240 m is still sufficiently high to resolve the majority of thin cirrus clouds. We note, that after April 2011, only daytime observations of DARDAR exist due to a battery anomaly of CloudSat.

In addition to the SEVIRI input, IceCloudNet is provided with static metadata detailing the location and time of an observation.
Besides the observation location, the use of latitude and longitude coordinates also conveys details of the spatial resolution of SEVIRI at a specific location, as resolution decreases with increasing distance from the nadir viewpoint at 0°E/0°N. 


We chose a domain from 30°W to 30°E and 30°S to 30°N, resulting in 1984$\times$1792 pixels per SEVIRI image in its native resolution. The reason for choosing this domain is two-fold: on the one hand, the resolution of SEVIRI is still high, even at the edges of the domain, given its proximity to the nadir viewpoint with an average horizontal resolution of 10.4 km\textsuperscript{2}; on the other hand, the tropics are an interesting area given the high uncertainties associated with the development of convective clouds and anvil cirrus \citep{gasparini_opinion_2023}. In the vertical dimension, we limit the dataset to altitudes between 4000 m and 17000 m resulting in 55 equally spaced vertical levels at a resolution of 240 m. This covers the whole range of cloud ice occurrence for the given latitudes.

For training and validating the neural network, we create patches with DARDAR overpasses at its center with a size of 256$\times$256 pixels, which corresponds to 768 km x 768 km, resulting in 89,229 patches for the years 2007-2012. This patch size strikes a balance between capturing cloud scenes from small to mesoscale systems in the same patch while still being computationally feasible. Formally, our dataset is described as:

\begin{itemize}
    \item $X \in \mathbb{R}^{89229 \times 256 \times 256 \times 11}$ for the eight \ac{ir} and three visible channels of SEVIRI.
    \item $X_{\textrm{meta}} \in \mathbb{Z}^{89229 \times 5}$ for latitude, longitude, month, the binary day/night mask, and the land/water mask (\textit{coast}, \textit{sea}, \textit{land}).
    \item $Y_{\textrm{masked}} \in \mathbb{R}^{89229 \times 256 \times 256 \times 55 \times 2}$ for \ac{iwc} and \ac{nice} DARDAR retrievals. Note, that while being on a 256$\times$256 grid, only a small subset of the grid points contain data, i.e. along the DARDAR overpass, which is represented as the variable $M$.
    \item $M = \mathbb{Z}_{2}^{89229 \times 256 \times 256 \times 1}$ for the binary overpass mask indicating the location of the satellite overpass.
\end{itemize}

Handling sunglint is a common challenge for satellite retrievals, as it can introduce artifacts over open water in the visible channels. However, due to the small viewing angles of SEVIRI in the region of interest, only limited areas are affected by sunglint. Furthermore, since \ac{ml} models have demonstrated the ability to distinguish between clouds and sunglint based on their differing spatial patterns, we do not explicitly address sunglint in this study \citep{schroder_generating_2002}.



\subsection{Inference dataset}\label{sec:data_inference}

Once trained and validated on the co-located SEVIRI and DARDAR data that is described in the previous section, IceCloudNet is able to predict vertical cloud profiles and their corresponding \ac{iwc} and \ac{nice} properties for all available SEVIRI observations in the domain that the model was trained on, i.e. independent of the availability of DARDAR overpasses. Complementing this paper are IceCloudNet predictions for \ac{iwc} and \ac{nice} for the years 2007 - 2017 on the native SEVIRI horizontal resolution, 240 m vertical resolution, and 15-minute temporal resolution for the domain from 30°W to 30°E and 30°S to 30°N (\url{https://www.wdc-climate.de/ui/entry?acronym=IceCloudNet_3Drecon}). The trained IceCloudNet model can be used to create custom predictions outside this time range as well. The predictions are made by first splitting a SEVIRI image into quadratic patches of 256$\times$256 pixel, followed by predicting the vertical profiles, and in the last step stitching the patches back together to the original image. The produced dataset increases the availability of vertical cloud profiles for years where DARDAR data are available by a factor of $1.8\times10\textsuperscript{6}$ and especially enhances the ability to study the temporal evolution of clouds by reducing the revisit time of 16 days by DARDAR to 15 minutes. The DARDAR data set is now discontinued, since the underlying satellite missions are not operational anymore and it also contains large gaps, caused by the CloudSat battery anomaly. For these periods, IceCloudNet enables an unprecedented opportunity to obtain vertical cloud profiles containing information about \ac{iwc} and \ac{nice} from geostationary satellite data.

\section{Methodology}\label{sec:methods}

\subsection{Problem setting}
The outlined task can be formally described as learning a mapping $f: \mathcal{X} = \mathbb{R}^{256 \times 256 \times 11} \rightarrow \mathcal{Y} = \mathbb{R}^{256 \times 256 \times 55 \times 2}$, where $x \in \mathcal{X}$ represents SEVIRI channels and $y \in \mathcal{Y}$ the \ac{iwc} and \ac{nice} of clouds containing ice. Note that at training time, a prediction $\hat{y} \in \mathcal{Y}$ is masked to a narrow swath $\hat{y}_{\textrm{masked}}$ using the corresponding overpass mask $m \in M$

    \begin{align*}
        \hat{y}_{\textrm{masked}} = m\odot\hat{y}
    \end{align*}

and the model is optimized on the masked data only. Despite the high level of sparsity in the ground truth data, IceCloudNet is capable of learning to translate SEVIRI input into vertical cloud profile across the full spatial coverage from only partial observations of the output.


\subsection{IceCloudNet architecture}\label{sec:methods_architecture}

\begin{figure}[h]
  \centerline{\includegraphics[width=39pc]{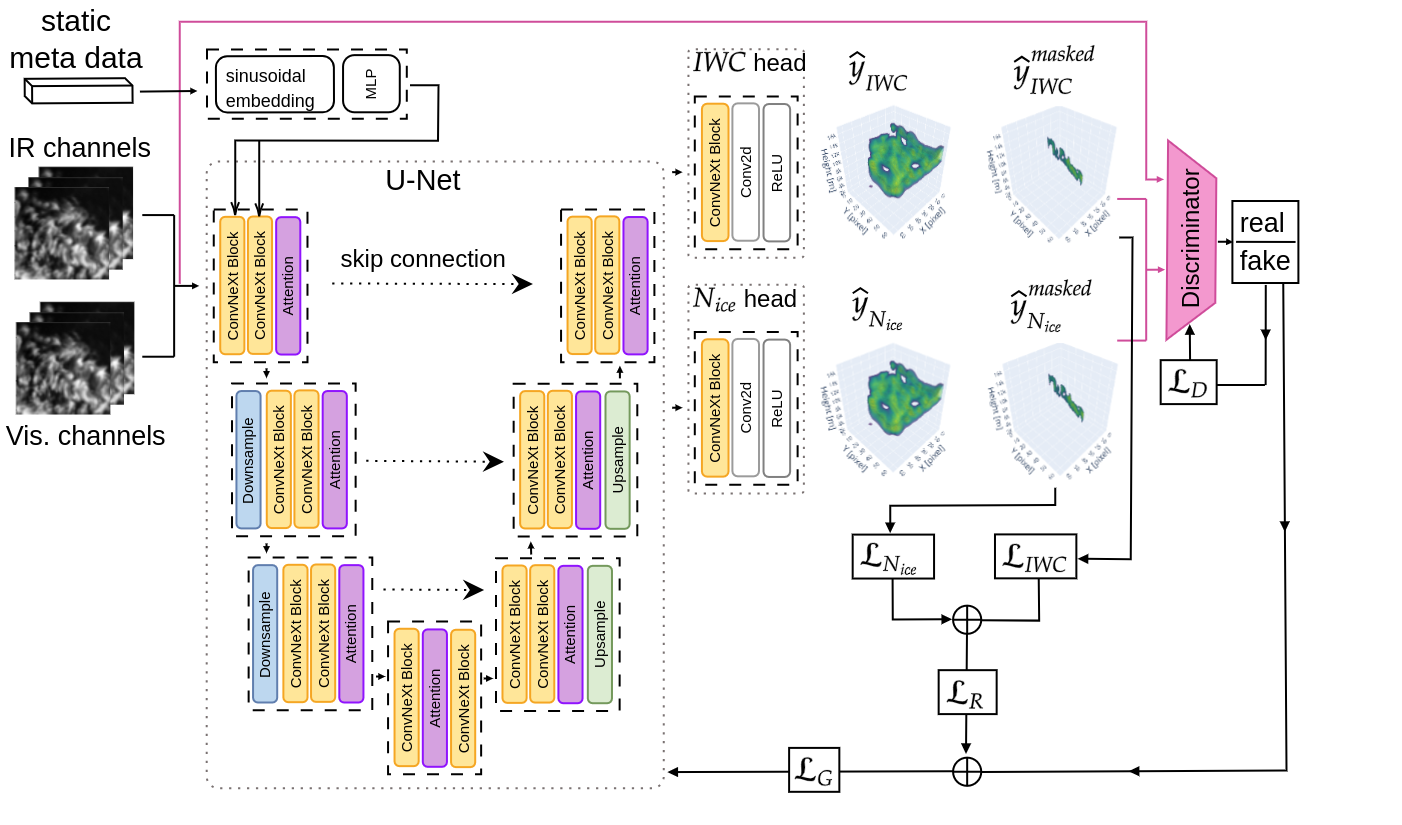}}
  \caption{A visualization of the IceCloudNet model architecture, including sample inputs and predictions, as well as the flow of losses during optimization. Abbreviations and acronyms used in the figure: Vis. - Visible; MLP - Multi Layer Perceptron; Conv2d - 2d convolution; ReLU - Rectified Linear Unit; $\mathcal{L}_{D}$ - Discriminator Loss; $\mathcal{L}_{R}$ - Regression loss; $\mathcal{L}_{G}$ - Loss of the main model generating cloud profiles.}
  \label{fig:architecture}
\end{figure}

The IceCloudNet model architecture consists of three main building blocks: A meta-data embedder $\mathnormal{E}$ that encodes static information such as location and temporal information (\ref{sec:architecture_meta}), a ConvNeXt-based U-Net $\mathnormal{U}$ which takes SEVIRI patches as input and generates three-dimensional cubes of \ac{iwc} and \ac{nice} via two prediction heads (\ref{sec:architecture_unet}), and a 3D-\ac{cnn}-based discriminator $\mathnormal{D}$ model acting as a learned similarity measure aimed at reducing the commonly observed blurring effect of models trained on pixel-based losses and thus leading to more accurate predictions (\ref{sec:architecture_disc}). IceCloudNet architecture and the losses used to train it are visualized in Fig. \ref{fig:architecture}.

\subsubsection{Meta data embedding}\label{sec:architecture_meta}

Cloud properties are subject to diurnal and seasonal cycles, that may depend on the location and Earth's surface type. Thus, we use a meta-data embedder $\mathnormal{E}$ that conditions predictions on the static information $x_{\textrm{meta}}$ for each patch. 


The meta-data embedding module $\mathnormal{E}$ works by following a similar approach as the time-step conditioning in the U-Net architectures commonly used in denoising diffusion frameworks~\citep[e.g.][]{ho2020denoising}. As shown in Fig.~\ref{fig:architecture}, the static metadata $x_{meta}$ is first subject to a sinusoidal embedding~\citep{vaswani2017,ho2020denoising}, followed by a learned embedding \ac{mlp}. The output of the \ac{mlp} is the shared embedding $emb$, which is injected into individual ConvNeXt encoding blocks. Since at each level of the U-Net, the internal feature representations have different numbers of channels, each ConvNeXt block is equipped with an auxiliary layer, which maps from the channels of the global $emb$ representation to the number of channels processed by that specific block:


    \begin{align*}
        emb = E(x_{\textrm{meta}})
    \end{align*}

\subsubsection{ConvNeXt U-Net}\label{sec:architecture_unet}

The backbone of IceCloudNet follows a U-Net~\citep{ronneberger_u-net_2015} topology (encoder-decoder architecture with additional skip connections at different downsampling levels), similar to the models used in recent work on denoising diffusion models~\citep{bansal2023,ho2020denoising}. More precisely, the exact topology is derived from the one used in the seminal \ac{ddpm} paper~\citep{ho2020denoising}, and, like in the implementation of cold diffusion~\citep{bansal2023}, using ConvNeXt blocks~\citep{liu_convnet_2022}, which have been shown to outperform traditional ResNet blocks~\citep{he_deep_2015}. ConvNeXt blocks use a larger kernel size in the first layer, increase the number of hidden channels considerably, and replace all \ac{relu} activations with a single \ac{gelu} placed before the last layer. Each upsampling and downsampling module consists of 2 ConvNeXt blocks, followed by self-attention. 


The main U-Net backbone is hence used to model the transformation function $U$ from the input SEVIRI channels $x$, conditioned on $x_{\textrm{meta}}$, to the target $y$ related to the cloud ice properties of interest:


    \begin{align*}
        \hat{y} = U(x, E(x_{\textrm{meta}}))
    \end{align*}

The final 3D cloud profile predictions $\hat{y}$ for \ac{iwc} and \ac{nice} are derived via two identical prediction heads, each consisting of a single ConvNeXt block followed by a Conv2d layer with kernel size 1 and a \ac{relu} activation (which forces the predicted output to be non-negative, as is desired for these physical properties). The 3D output is achieved by treating the channel dimension of the final Conv2d as the vertical dimension.

The experimentation process involved tests with 3D convolutions in the decoding arm of the U-Net, but superior results were achieved with the described approach at a lower computational cost. We use the mean absolute error ($\ell_1$) as a metric to assess the similarity between $\hat{y}_{masked}$ and $y_{masked}$ during training, resulting in the following regression loss $\mathcal{L}_{R}$:

    \begin{align*}
        \mathcal{L}_{R} = \mathcal{L}_{N_{\textrm{ice}}} + \mathcal{L}_{\textrm{IWC}}
        = \ell_1\left(m\odot\hat{y}^{\textrm{IWC}} ,m\odot y^{\textrm{IWC}}\right) +\ \ell_1\left(m\odot\hat{y}^{N_{\textrm{ice}}} ,m\odot y^{N_{\textrm{ice}}}\right)
    \end{align*}

While $\ell_1$ produces less blurring compared to using a root mean squared error ($\ell_2$), it is widely acknowledged that pixel-based losses, especially for \ac{cnn}-based architectures, produce blurry outputs~\citep{larsen_autoencoding_2016}, which motivates the extension of the objective to include an adversarial term as described below.

\subsubsection{Discriminator}\label{sec:architecture_disc}

Apart from the $\ell_1$ objective, we incorporate an adversarial training procedure with a patch-based discriminator $\mathnormal{D}$ \citep{isola_image--image_2018} that is tasked to distinguish real cloud profiles (DARDAR) from reconstructed cloud profiles (U-Net predictions). The discriminator model acts as a learned metric that evaluates the similarity between predictions and reference data. The discriminator $\mathnormal{D}$ employed here is based on a PatchGAN \citep{isola_image--image_2018} architecture, expanded to 3D convolutions and is conditioned on SEVIRI images. This idea is inspired by VAE-GAN \citep{larsen_autoencoding_2016} where a \ac{vae} functioning as an image generator produces images of higher quality when trained in an adversarial fashion. A similar approach was applied in a 3D setting by \citet{buhmann_getting_2021} to simulate sparse particle showers.

The discriminator model $D$ is optimized with a Wasserstein loss \citep{arjovsky_wasserstein_2017} where

    \begin{align*}
        \mathcal{L}_{D} = D(x ,m\odot \hat{y}) - D(x,m\odot y)
    \end{align*}

is the minimisation objective for $D$ during training and the main model is optimized with the loss of $\mathcal{L}_{G}$ (an extended version of the regression loss $\mathcal{L}_R$):
    
    \begin{align*}
        \mathcal{L}_{G} = (\mathcal{L}_{N_{\textrm{ice}}} + \mathcal{L}_{\textrm{IWC}}) - \lambda D(x,m\odot \hat{y})
    \end{align*}
    
We follow the adaptive weighting approach used in \citet{esser_taming_2021} to dynamically adjust the discriminator weight $\lambda$. IceCloudNet is trained end-to-end where $U$ minimizes $\mathcal{L}_{G}$ and $D$ minimises $\mathcal{L}_{D}$. To create 3D cloud profiles during inference, only $E$ and $U$ are needed:
    
    \begin{align*}
        \hat{y} = U(x, E(x_\textrm{meta}))
    \end{align*}


\subsection{Experimental setup}

Data for the year 2010 is used as an independent test set. The remaining years are used for training, where 10 \% of patches are used as validation data for evaluating architectural choices and hyperparameters. In order to prevent spatiotemporal autocorrelation in training and validation splits, data from the same day will be assigned to only one split. Patches without any cloud ice in the DARDAR overpass are not used for training along with patches containing missing SEVIRI data. At test time, all patches are evaluated.

Since the target variables span multiple orders of magnitude, they are logarithmically ($log(\epsilon+x)$, with $\epsilon=1$) transformed. Before the logarithmic transformation \ac{iwc} values are scaled by 10\textsuperscript{7} and \ac{nice} by 10\textsuperscript{-2}.
SEVIRI channels are normalized by their mean and standard deviation calculated on the training dataset. SEVIRI's visible channels are only available during the day, they are hence masked for nighttime predictions.

We apply a random number of 90° rotations (with uniform probability) for data augmentation and train IceCloudNet for 100 epochs and batch size 64 using the Adam optimizer and learning rate of 10\textsuperscript{-6}. For regularization stochastic depth dropout \citep{huang_deep_2016} with a dropout probability of 25 \% is applied.

Due to the size of the network and dataset, a full-scale hyperparameter search for the model architecture is not computationally feasible. IceCloudNet consists of 25.1 million parameters in total, of which 15 million are for the U-Net, 11 million for the PatchGAN discriminator and 0.1 million for the positional embedding. For each component, an experiment with a decreased and increased number of parameters was conducted. The chosen setup outperformed the downsized model and is on-par with the larger model. 


\section{Results and Evaluation}\label{sec:results}

In this section, IceCloudNet is thoroughly evaluated and the strengths and shortcomings of its predictions are presented in detail. The evaluation of IceCloudNet poses two main challenges: First, the sparse availability of DARDAR reference data constrains the quantitative evaluation of the predictions to locations covered by the satellite overpass. Second, pixel-based evaluation metrics, which are commonly employed for \ac{ml} model evaluation, are not completely aligned with the quality of the predictions. To illustrate this, a prediction of a thin cloud shifted by one height level should intuitively be considered a better prediction than not predicting that cloud at all. However, a pixel-based loss would result in a smaller error for the prediction with no cloud present than for the prediction of the shifted cloud.


To tackle these challenges, IceCloudNet predictions are evaluated in the following along three spatial abstraction levels:

\begin{enumerate}
    \item \textbf{Pixel-based:} Metrics calculated per pixel for IceCloudNet predictions against DARDAR reference data. This is only possible along the satellite overpasses. One pixel represents a single data point in a satellite overpass with a dimension of 3 km x 3 km x 240 m.
    \item \textbf{Overpass-based:} Averaged values or engineered metrics along the overpass, such as zonal means or cloud cover along the overpass.
    \item \textbf{Full domain:} Here, no DARDAR reference data are available, and we are limited to evaluating the distribution statistics of IceCloudNet predictions. By using a whole year of data for evaluation, we can assume that the predictions of the whole spatial domain of IceCloudNet should follow a distribution similar to DARDAR overpasses, thus providing a sanity check for our predictions. 
\end{enumerate}

In the following, we first assess the ability of IceCloudNet to correctly represent the occurrence of cloud ice in the three-dimensional space (subsection \ref{sec:results}\ref{sec:results_cloud_occurrence}) and second, its ability to correctly determine the magnitude of cloud ice, namely \ac{iwc} and \ac{nice} (subsection \ref{sec:results}\ref{sec:results_cloud_ice}). All evaluations shown in this section have been conducted on the year 2010, which IceCloudNet had no exposure to during training, i.e. the independent test dataset. 
In subsection \ref{sec:results}\ref{sec:case_studies})IceCloudNet predictions are compared to ground-based observations in two representative case studies.

\begin{table}[t]
\caption{Pixel-based IceCloudNet performance on independent test dataset (2010). Note that all metrics can only be calculated where DARDAR ground-truth data are available. The classification metrics accuracy (Acc), precision (Pr), and recall (Re) are calculated on a post-processed cloud mask where a pixel contains a cloud if the \ac{iwc} $>$ 10\textsuperscript{-7} kg m\textsuperscript{-3}. $R^2$ is calculated on logarithmically transformed target variables. The mean absolute error (MAE) is calculated as an in-cloud error on the original scale. Arrows illustrate if smaller or larger values of a metric indicate better performance.}
\label{table:pixel_results}
\centering
\begin{tabular}{@{}lcccccccc@{}}
\toprule
& \textbf{Time of day}  & \ac{iwc} \textbf{MAE} [kg m\textsuperscript{-3}]  $\downarrow$ & \ac{nice} \textbf{MAE} [m\textsuperscript{-3}]  $\downarrow$  & \ac{iwc} $\mathbf{R^2}$ $\uparrow$ & \ac{nice} $\mathbf{R^2}$ $\uparrow$ & \textbf{ Acc (\%)} $\uparrow$ & \textbf{ Pr (\%)} $\uparrow$ & \textbf{ Re (\%)} $\uparrow$ \\ \midrule
      & all day      & 5.6$\times$10\textsuperscript{-5} & 6.5$\times$10$^4$  & 0.69 & 0.54 & 96 & 78 & 75  \\
      & night time   & 4.4$\times$10\textsuperscript{-5} & 5.6$\times$10$^4$ & 0.70 & 0.54 & 95 & 79 & 74 \\
      & daytime      & 7.5$\times$10\textsuperscript{-5} & 8.0$\times$10$^4$ & 0.69 & 0.54 & 97 & 77 & 75 \\ \bottomrule
\end{tabular}
\end{table}


\subsection{Cloud occurence evaluation}\label{sec:results_cloud_occurrence}

Pixel-based performance on the independent test set is reported in Table~\ref{table:pixel_results}. IceCloudNet is able to achieve an accuracy of 96 \% in predicting which pixels contain a cloud and which do not. Since 92 \%  of the pixels in our dataset do not contain a cloud, considering the accuracy itself does not provide a balanced perspective. Better suited for evaluating the ability to predict a cloud in a given pixel are precision and recall. Precision is a metric that measures the fraction of positive predictions (in this case, cloudy) that are also positive in the ground truth dataset. On the other hand, recall calculates the proportion of all positive cases in the ground truth dataset that were correctly predicted as positive. Even at the pixel level, IceCloudNet is able to predict the occurrence of clouds containing ice crystals with precision and recall of 78 \% and 75 \%, respectively.

It is worth noting that the performance for nighttime observation slightly outperforms daytime predictions, despite daytime predictions having additional input channels in the visible spectrum that provide additional information about cloud extent and optical thickness. We hypothesize that the worse performance during the day is due to solar radiation that interferes with the lidar instrument of CALIPSO introducing higher retrieval uncertainties \citep{hunt_calipso_2009}, meaning that there could be a compensating effect between additional information from visible SEVIRI channels and deteriorated quality of DARDAR retrievals during the day.


In this work, we forego the comparison with simpler models which do not capitalize on spatial dependencies in the input data, since earlier works have shown that CNN-based approaches outperform those simpler model architectures \citep{amell_ice_2022, jeggle_icecloudnet_2023}. 

\begin{figure}[h]
  \centerline{\includegraphics[width=39pc]{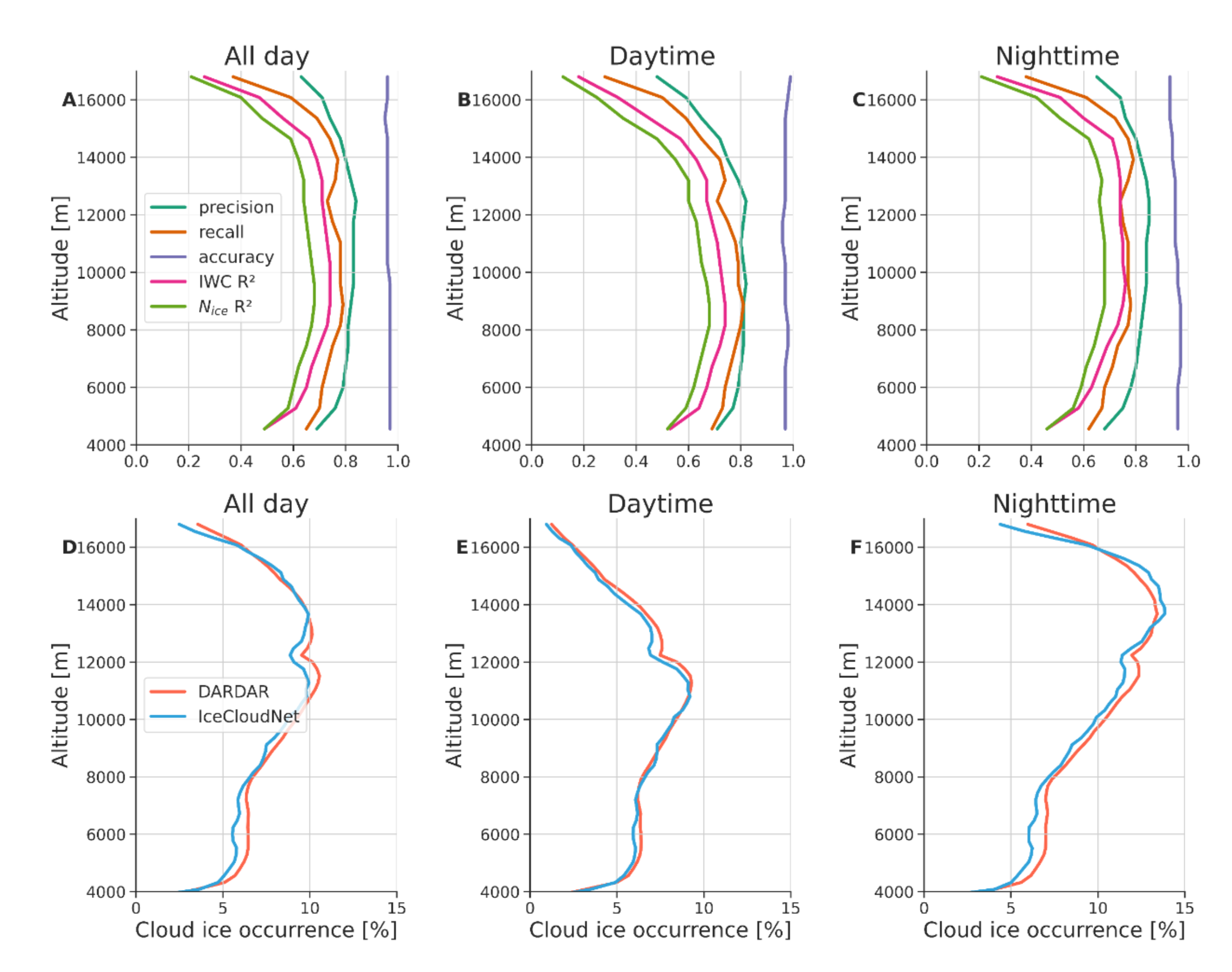}}
  \caption{Regression and classification metrics calculated per pixel of all (A), daytime (B), nighttime (C) DARDAR overpasses for different height levels for 2010. Each height level in (A),(B), and (C) consists of three height levels in the original resolution (240 m). The classification metrics are calculated on a post-processed cloud mask and $R^2$ is calculated on logarithmically transformed \ac{iwc} and \ac{nice}. All day (D)/ daytime (E) / nighttime (F) cloud occurrence for different height levels for DARDAR overpasses (red), and IceCloudNet predictions along DARDAR overpasses (blue)}
  \label{fig:cloud_occurrence}
\end{figure}


Since IceCloudNet predictions are vertically resolved, we are interested in its ability to represent cloud ice occurrence for different height levels and if its performance differs depending on the height. Fig.~\ref{fig:cloud_occurrence}~A - C display the classification metrics discussed previously as a function of altitude for all data, daytime data, and nighttime data, respectively. The skill to correctly predict the occurrence of cloud ice is highest between 5 km and 14 km. For thin cirrus occurring above 15 km the precision and recall metrics drop slightly, yet still 70 \% of predicted cloud pixels are correctly predicted at its minimum performance. We assume that the main reason for the worsening of the performance at high altitudes is that passive imagers such as SEVIRI are known to have a limited sensitivity to thin cirrus clouds~\citep[e.g.][]{holz_global_2008,ackerman_cloud_2008}, which dominate at these heights~\citep{kramer_microphysics_2020,gasparini_opinion_2023}. IceCloudNet also struggles to accurately represent clouds at heights below 6000 m. We assume that this is mainly due to the fact that those lower clouds are often masked by higher cloud layers in the underlying SEVIRI data. For daytime data (B), where visible channels are available for IceCloudNet predictions, recall at lower altitudes is better. Visible channels are sensitive to cloud albedo, which is typically higher for clouds at lower altitudes, explaining the increased performance at lower altitudes. The opposite effect is observed for nighttime data (C), where the recall curve declines more for altitudes below 8 km. Another potential cause for the decreased performance at low altitudes could be due to the inherent uncertainties behind the cloud phase retrieval in DARDAR since both liquid and ice phases can occur at these heights. For high altitudes, on the contrary, it can be observed that both recall and precision have higher values for nighttime data. Since high cirrus clouds are typically semi-transparent to solar radiation, visible channels provide less added value, but the increased retrieval uncertainty of DARDAR data due to solar radiation is affecting the data quality of the reference data during training and hence the prediction performance.

Figures~\ref{fig:cloud_occurrence}~D - F show the occurrence of clouds containing ice as a function of altitude in the test dataset for IceCloudNet and DARDAR reference data along the satellite overpasses. IceCloudNet is able to reproduce the shape of the height-dependent cloud occurrence curve, but tends to underestimate cloud occurrence on avarage by 0.4 \% with the largest deviations for altitudes for altitudes between 4 km and 8 km, around 12 km and above 16 km. For daytime data (E), IceCloudNet almost perfectly reproduces cloud occurrence with a mean deviation of 0.2 \%. During the night, when no visible channels are available, IceCloudNet underestimates cloud occurrence up to 8 km by 0.6 \% on average. Its performance is best between 8 km and 15 km, before the deviation from the DARDAR reference data increases again to 1.5 \% at 17 km. At these altitudes, mainly optically thin \ac{ttl} cirrus occur, which the SEVIRI instrument is less sensitive to. 




\begin{figure}[h]
  \centerline{\includegraphics[width=39pc]{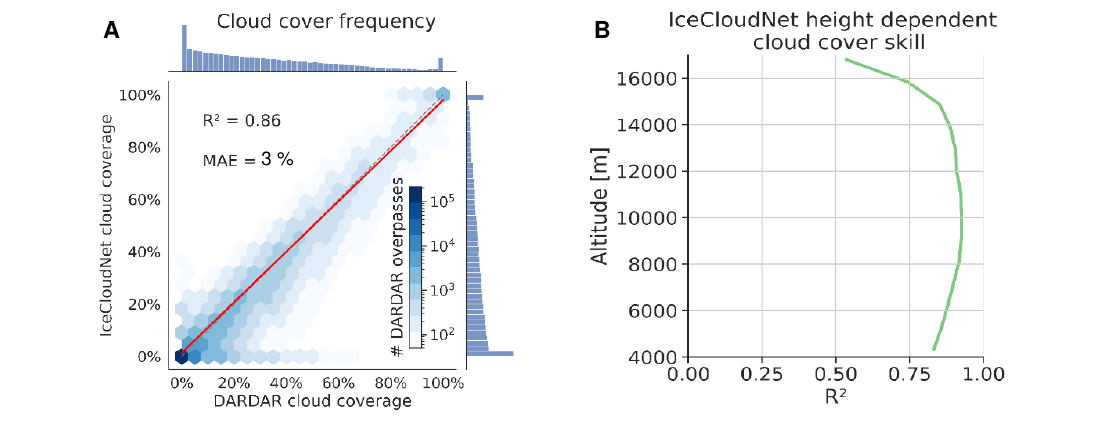}}
  \caption{Cloud cover along DARDAR overpasses: (A) Frequency of cloud cover along DARDAR overpass profiles for DARDAR reference data and IceCloudNet predictions. A single data point here represents the cloud cover along the DARDAR overpass in a single patch ( 256$\times$256 pixels ) as used in the training of IceCloudNet. The red line shows the linear regression fitted between the predictions and ground truth data. The marginal distributions of cloud cover frequencies are shown on the right and top, respectively. Cloud cover is calculated such that an atmospheric column is cloudy as soon as there is a cloudy pixel at any height, for example, if all columns within an overpass profile contain at least one cloudy pixel, the cloud cover is considered 100 \%. (B): The coefficient of determination ($R^2$) calculated for different height levels between DARDAR and IceCloudNet cloud cover for all overpass profiles of the test dataset. A height level bin in this figure corresponds to 3 levels of the actual data, that is, 720 m.}
  \label{fig:cloud_cover}
\end{figure}

While examining pixel-based metrics is crucial for understanding the performance of IceCloudNet, it is also valuable to consider measures beyond these metrics. 
For instance, the comparison between predicted versus actual cloud cover along the DARDAR overpass enables are more holistic evaluation of a cloud scene.
Here, cloud cover is calculated such that an atmospheric column is cloudy if there is at least one cloudy pixel at any height. Fig. \ref{fig:cloud_cover} A displays IceCloudNet vs. DARDAR cloud cover predictions and their absolute frequencies in the test dataset, where one data point is the cloud cover in a satellite overpass for a single 256$\times$256 pixel patch. The length of the satellite overpass in a patch depends on the angle of the overpass, with a mean length of 1$\times$288 pixels. 

IceCloudNet is able to achieve an $R^2$ of 0.86 and mean absolute error of 3~\% in predicting cloud cover along the satellite overpass. This means that most cloud scenes are correctly represented by IceCloudNet. Fig. \ref{fig:cloud_cover} B allows for a more detailed view of the predicted cloud cover performance for different height levels. The plot shows a high $R^2$ of $>$ 0.8 for levels between 5 km and 15 km, with a maximum of 0.9 between 10 km and 12 km. Like in Fig. \ref{fig:cloud_occurrence}, the values drop above and below this range. Up to 17 km, $R^2$ stays above 0.5, indicating that IceCloudNet may not predict the exact location of a cloudy pixel at these heights, but it is still capable of representing the overall cloud scene. 

\subsection{Cloud ice evaluation}\label{sec:results_cloud_ice}

In addition to correctly representing the occurrence and macrophysical structure of clouds, it is critical to provide a good estimate of the microphysical properties \ac{iwc} and \ac{nice} along the entire atmospheric column. These properties are key to determining the radiative properties of clouds and are indicative of their interactions with aerosol particles.

\begin{figure}[h]
  \centerline{\includegraphics[width=\textwidth]{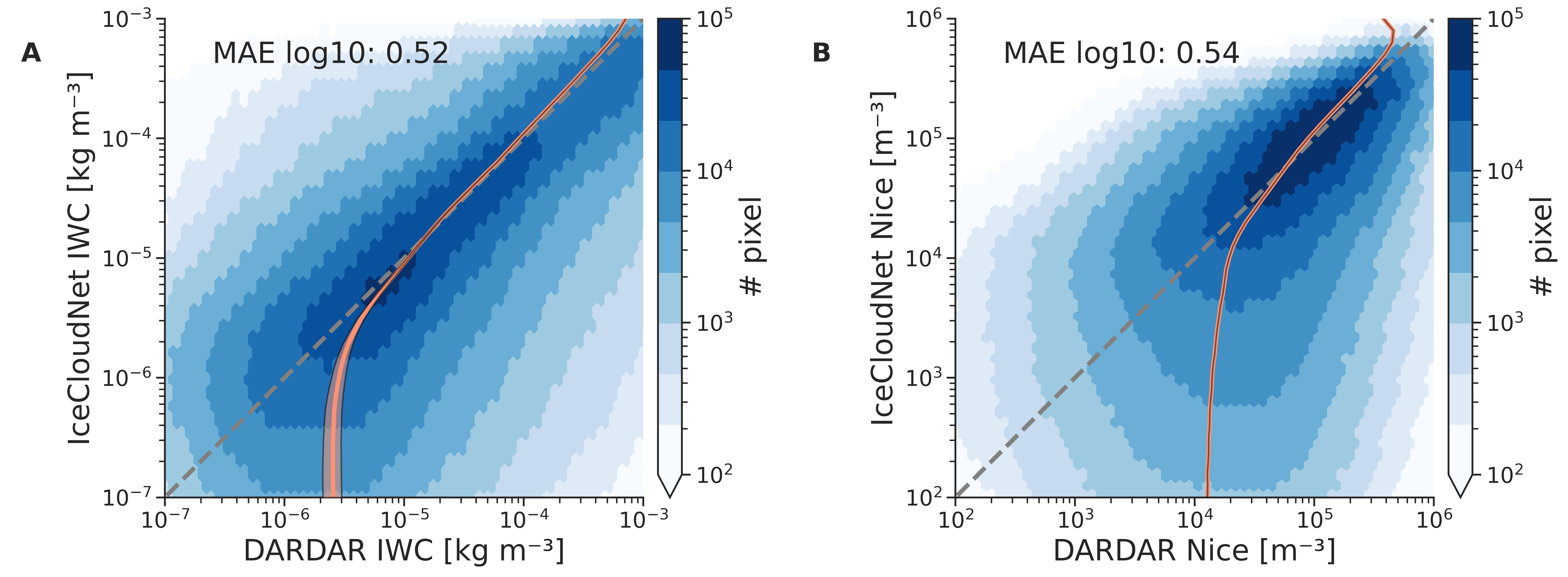}}
  \caption{Frequency of in-cloud DARDAR and IceCloudNet \ac{iwc} (\textbf{A}) and \ac{nice} (\textbf{B}) for the test dataset of 2010. Mean absolute errors (MAE) are calculated on in-cloud values and are shown on a log transformed scale. The orange line and shaded area denotes the mean and standard deviation of DARDAR values calculated for bins of IceCloudNet prediction values, with a bin size of $\frac{1}{10}$ of an order of magnitude.}
  \label{fig:cloud_ice_comparison}
\end{figure}

IceCloudNet predictions achieve an $R^2$ of 0.69 and 0.54 for \ac{iwc} and \ac{nice}, respectively (Table \ref{table:pixel_results}). Fig \ref{fig:cloud_ice_comparison} visualizes the reference in-cloud \ac{iwc} (A) and \ac{nice} (B) of DARDAR against the predictions of IceCloudNet. IceCloudNet generally tends to slightly underestimate \ac{iwc} and \ac{nice} with a mean absolute error in logarithmic space of 0.52 (\ac{iwc}) and 0.54 (\ac{nice}), meaning that on average IceCloudNet predictions are off by approximately half an order of magnitude. Fig. \ref{fig:cloud_ice_comparison} shows that the underestimation occurs at the lower end of the data distributions, that is, for \ac{iwc} $<$ 5$\times$10$^{-6}$ kg m$^{-3}$ and \ac{nice} $<$ 2$\times$10$^4$ m$^{-3}$. For \ac{iwc} and \ac{nice} 75 \% of DARDAR in-cloud data are above these thresholds. We assume that the \ac{iwc} prediction performance is outperforming \ac{nice} predictions due to stronger sensitivity of the cirrus radiative effects to \ac{iwc} than to \ac{nice}, and hence an increase in the signal in the SEVIRI data.

\begin{figure}[h]
  \centerline{\includegraphics[width=39pc]{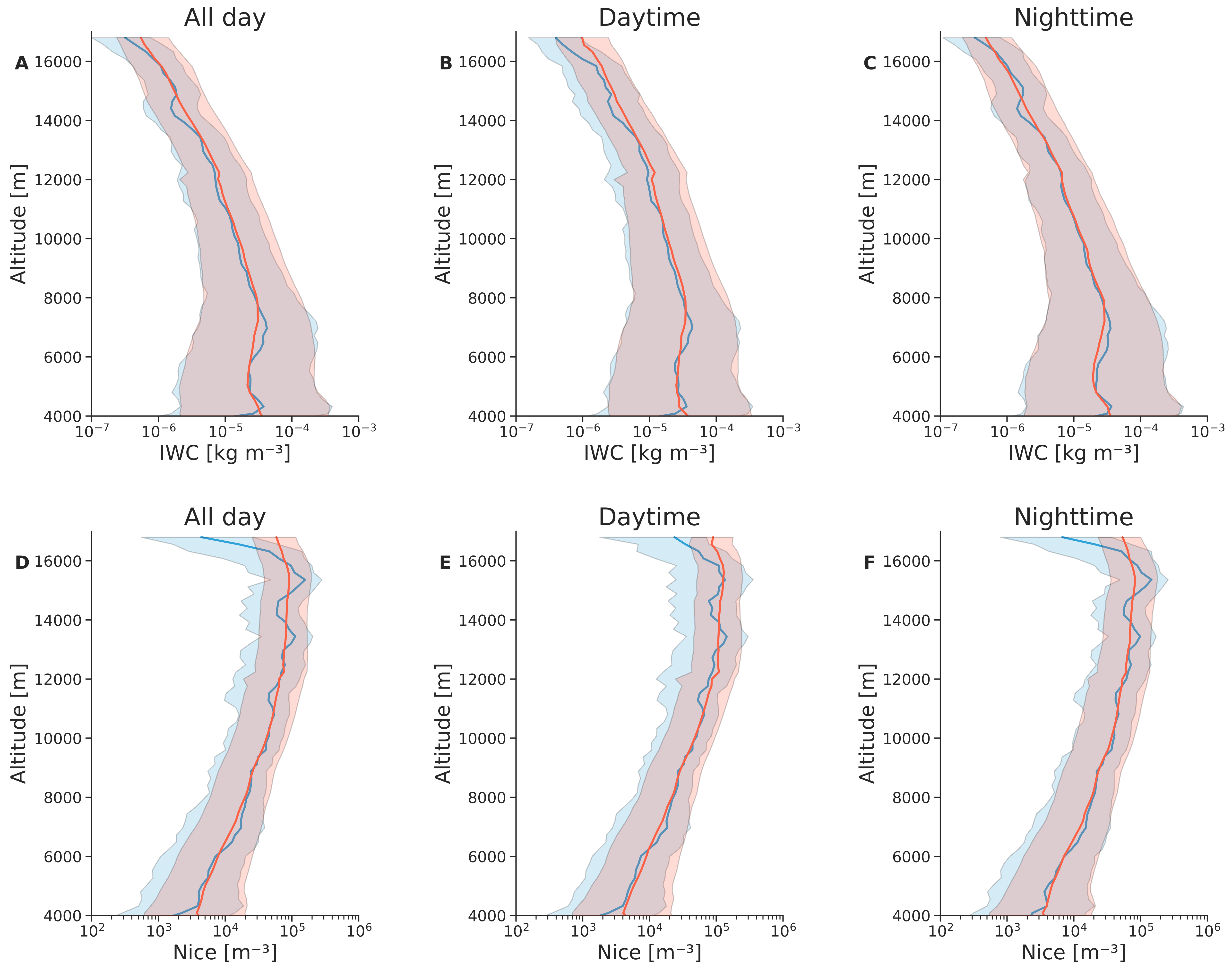}}
  \caption{Comparison of IceCloudNet (blue) and DARDAR (red) cloud ice properties for different height levels. Solid lines represent median values, and shaded areas represent the interquartile range. Panels A,B,C display \ac{iwc} values for all observations (A), daytime observations (B), and nighttime observations (C). Panels D,E,F analogously show values for \ac{nice}.}
  \label{fig:cloud_ice_height}
\end{figure}

Similar to the cloud occurrence performance (Fig. \ref{fig:cloud_occurrence}), the ability to correctly quantify microphysical properties is dependent on altitude and time of day. Fig. \ref{fig:cloud_ice_height} displays median values and interquartile ranges for \ac{iwc} and \ac{nice} for all-day, daytime, and nighttime data.  It can be seen that throughout the atmospheric column IceCloudNet closely reproduces  the median and quartile values for both \ac{iwc} (A) and \ac{nice} (D) with mean deviations of the median of 0.006 and 0.04 orders of magnitude, respectively. As for the cloud occurrence performance, the deviations are largest for high altitudes, i.e. $>$ 16 km during the night (C, F) and $>$ 15 km during the day (B, E). Given the decreased occurrence of high cirrus during the day (Fig. \ref{fig:cloud_occurrence}), the lower performance of IceCloudNet during the day at high altitudes only slightly impacts its overall performance. Despite the additional information from visible channels during the day, no performance improvement can be observed at lower altitudes for daytime predictions in the median and quartile values. When considering $R^2$ as a function of altitude (Fig. \ref{fig:cloud_occurrence} A,B,C), a slightly improved performance can be observed during the day for altitudes below 8 km.  



\begin{figure}[h]
  \centerline{\includegraphics[width=33pc]{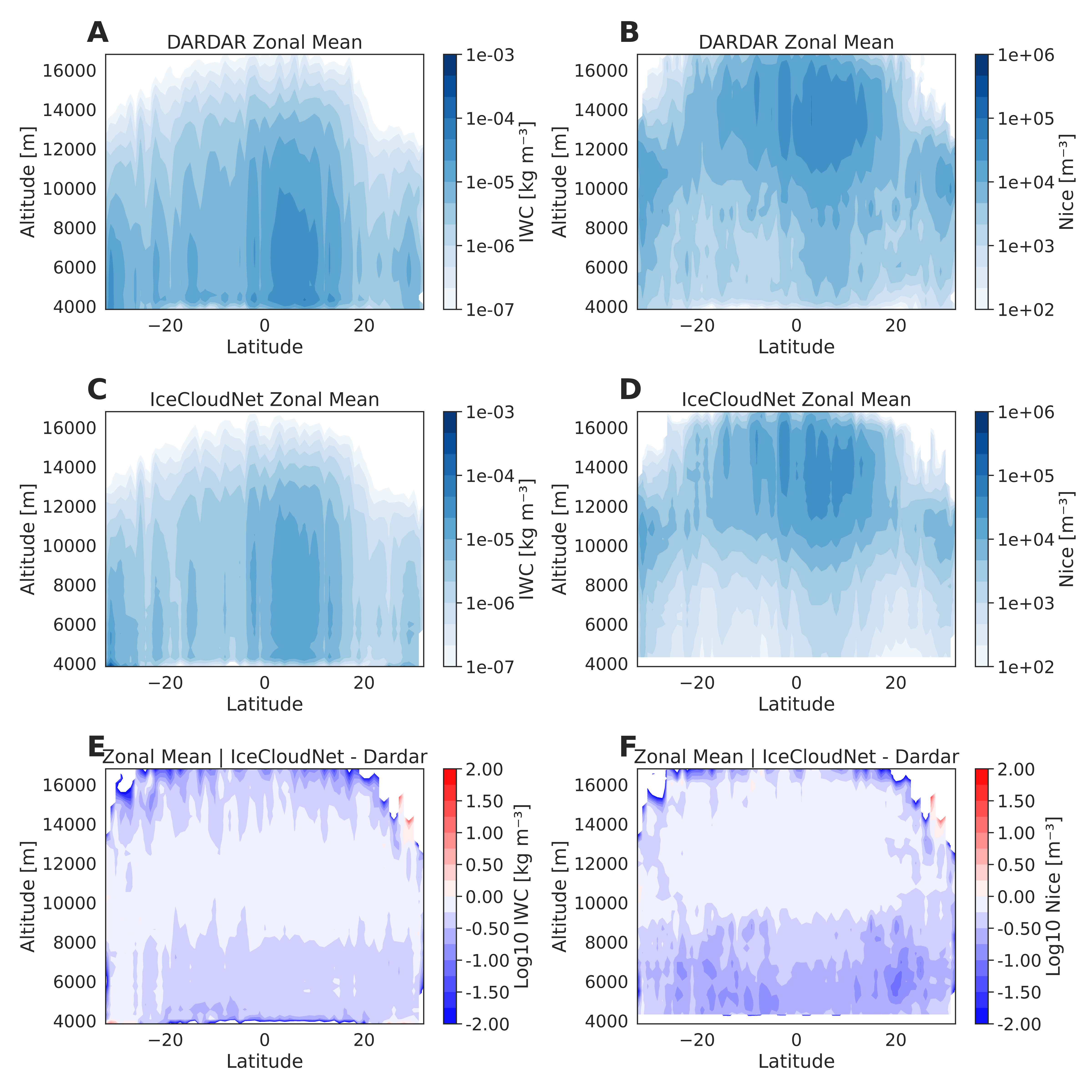}}
  \caption{Zonal \ac{iwc} grid mean values  for DARDAR \textbf{A}, IceCloudNet \textbf{C}, and the difference between both \textbf{E} for data along DARDAR overpasses. Panels \textbf{B}, \textbf{C}, and \textbf{F} analogously display zonal in-cloud means for \ac{nice}.}
  \label{fig:zonal_mean}
\end{figure}

Zonal mean plots provide a big picture view on IceCloudNet's performance for different latitudes over the area of interest. Fig. \ref{fig:zonal_mean} shows the zonal mean of DARDAR and IceCloudNet for \ac{iwc} and \ac{nice} along the satellite overpass, as well as the differences between both. It can be seen that IceCloudNet is able to reproduce the spatial patterns of DARDAR data, but underestimates \ac{iwc} and \ac{nice} across the whole spatial domain. Both, the vertical distribution of the microphysical properties as well as the zonal distribution is well captured by IceCloudNet. Consistent with the results shown above, the largest deviation from the reference data can be observed at altitudes below 8 km for \ac{iwc} (0.5 orders of magnitude) and 9 km for \ac{nice} (0.75 orders of magnitude), as well as for altitudes above 14 km. We note that the representation of zonal mean \ac{iwc} and \ac{nice} is not directly comparable here, given that the former is calculated as grid-mean and the latter as in-cloud mean as often done in climate models \citep[e.g.][]{lohmann_future_2020}.

\begin{figure}[h]
  \centerline{\includegraphics[width=39pc]{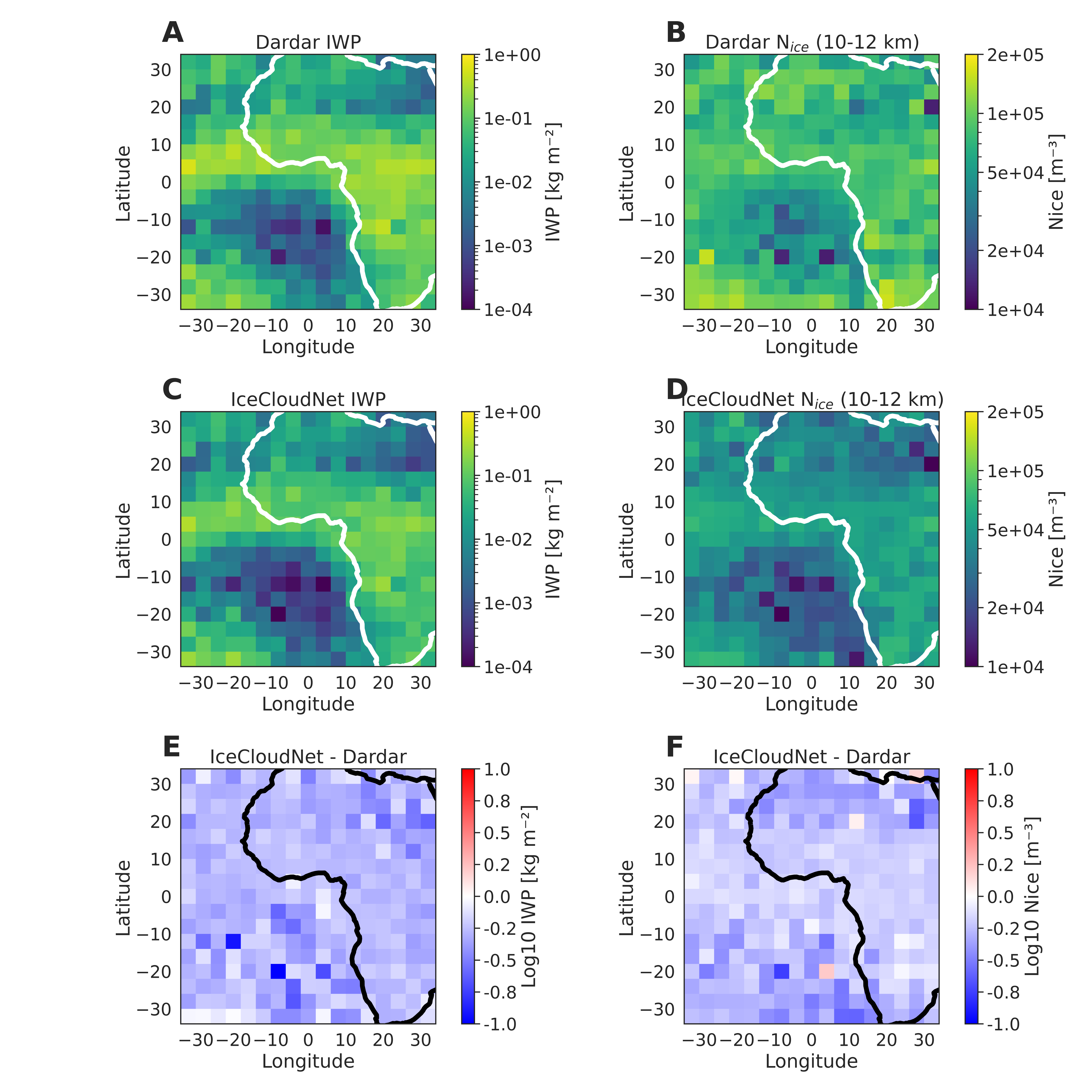}}
  \caption{Annual mean \ac{iwp} values of the whole spatio-temporal domain of the test dataset for DARDAR (A), IceCloudNet (C), and the difference between both (E) for data along the satellite overpasses. Panels (B), (D), and (F) analogously display \ac{nice} in-cloud mean values for altitudes between 10 km and 12 km. The data are aggregated to a grid size of 5°$\times$5° as DARDAR does not cover the whole area due to its narrow swaths. The \ac{iwp} is not directly predicted by IceCloudNet, but calculated based on its \ac{iwc} predictions.}
  \label{fig:iwp_grid_mean}
\end{figure}

In order to view the regional distribution of DARDAR and IceCloudNet, we calculate \ac{iwp} by integrating the \ac{iwc} predictions and average it onto 5°$\times$5° gridboxes. For \ac{nice} a vertical integration is not useful, we instead calculate the in-cloud mean for altitudes between 10 km and 12 km to analyses spatial patterns. Fig. \ref{fig:iwp_grid_mean} shows that IceCloudNet predictions are in good agreement with the reference data, especially in the \ac{itcz} where clouds containing ice occur most often. In areas with a lower cloud ice appearance, namely northern Africa and the Atlantic Ocean west of Africa, IceCloudNet performance decreases with a deviation in logarithmic space between 0.25 to 1 kg m\textsuperscript{-2} for \ac{iwp} and 0.25 to 1 m\textsuperscript{-3} for \ac{nice}, likely due to the lack of training data. 


\subsection{Case Studies}\label{sec:case_studies}

\begin{figure}[ht!]
  \centerline{\includegraphics[width=33pc]{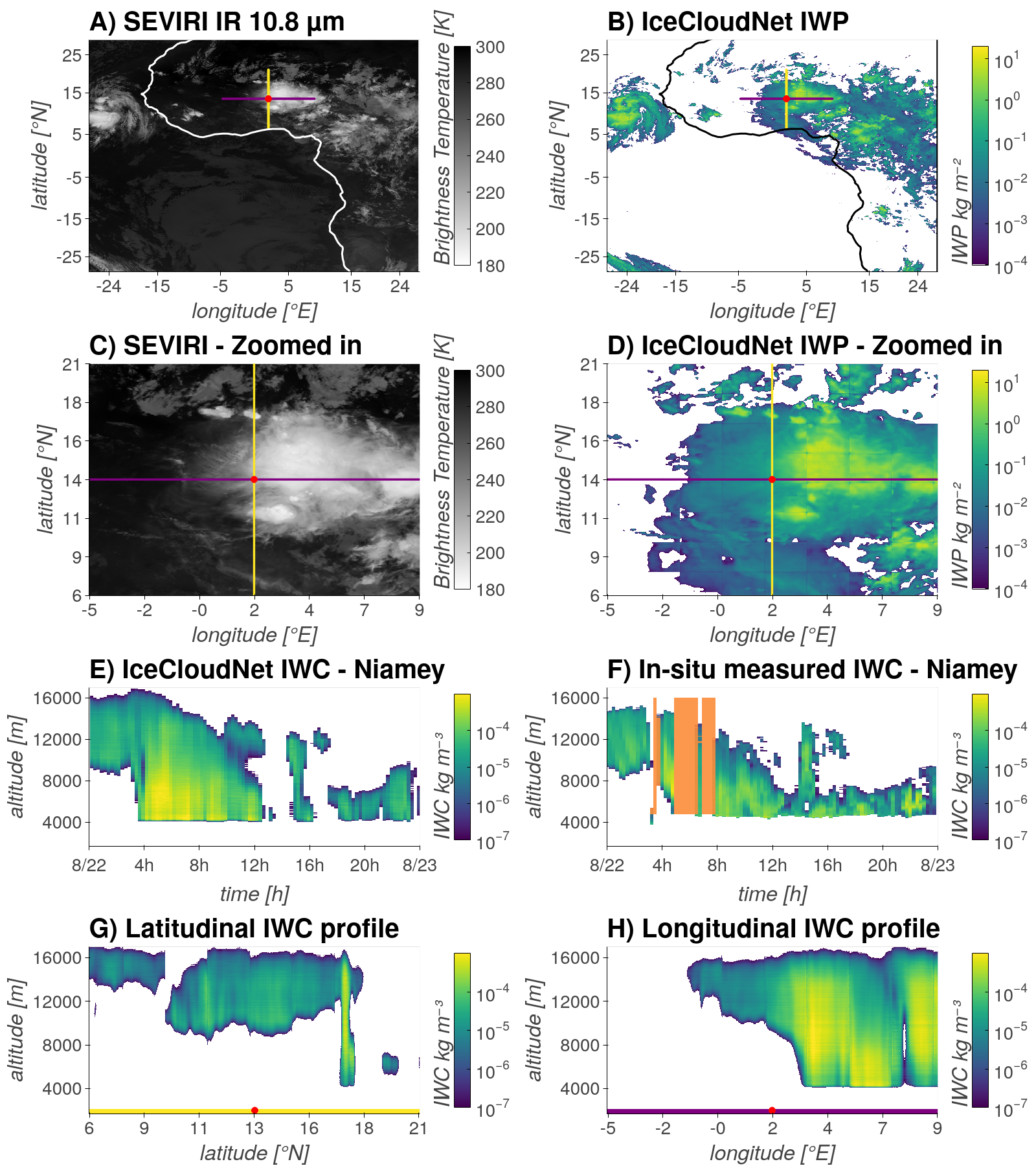}}
  \caption{Case study of IceCloudNet predictions for 2006-08-22 compared with ground based station in Niamey. The station data is averaged to 15 minute time intervals. Panel A shows one of the SEVIRI \ac{ir} input channels used for the IceCloudNet prediction, which is visualized with IWP predictions in panel B. Panels C and D are zoomed in maps of SEVIRI and IceCloudNet \ac{iwp} with Niamey at the center. IceCloudNet \ac{iwc} predictions at the location of the station in Niamey (red dot) for one full day are displayed in E and the actually retrieved \ac{iwc} at the Niamey station in panel F. Missing data for the station is shaded in orange. A latitudinal and longitudinal cross-section of IceCloudNet \ac{iwc} predictions, marked by the purple and yellow lines in A and B, are displayed in panels E and F, respectively. Note, that all panels except E and F show data for a single point in time, that is 2006-08-22 00:12:41.}
  \label{fig:case_study_niamey}
\end{figure}
\FloatBarrier

The key motivation for creating IceCloudNet was the lack of a temporally resolved perspective of clouds in 3D space. In the following, we show two case studies of IceCloudNet's ability to represent the evolution of cloud ice evaluated against observations made by ground-based stations of the Cloudnet \citep{illingworth_cloudnet_2007} and \ac{arm} \citep{jensen_atmospheric_2008} networks. The ground-based stations are equipped with radar and lidar instruments that retrieve \ac{iwc} measurements that we compare against. We resample the 30 s temporal resolution to the 15 min resolution of IceCloudNet and keep its original vertical resolution ranging from 29 to 51 m depending on the height level. The dates of both case studies have been randomly selected from the available data of the corresponding stations. IceCloudNet predictions come at a $\sim$3$\times$3 km$^2$ resolution, which is considerably coarser than the sampling area of the ground-based lidar and radar instruments, hence we would not expect an exact agreement between both data sources, even if the IceCloudNet predictions were perfect.

\subsubsection{Niamey - 2006-08-22}


In the first case study, IceCloudNet predictions for 2006-08-22 at the \ac{arm} station in Niamey, Niger are shown in Fig.~\ref{fig:case_study_niamey}. Panel A provides a holistic view of the cloud scene present at 00:12:41 via the 10.8 $\mu m$ channel of SEVIRI and panel C shows a zoomed in area around Niamey. A \ac{mcs} with multiple convective cores can be visually detected, which moves to the northwest throughout the day (not shown). 


By visually evaluating the predicted \ac{iwp} in Fig.~\ref{fig:case_study_niamey} B and D, it can be seen that all cloud structures apparent in A and C are well represented, in addition to thin cirrus clouds that are difficult to visually detect in the SEVIRI IR channel shown in Fig~\ref{fig:case_study_niamey} A and C. Panels E and F showcase IceCloudNet \ac{iwc} predictions and ground-based \ac{iwc} observations at Niamey for 24 hours. IceCloudNet is able to represent cloud shape and extent throughout the course of the day with a mean deviation of cloud top height of 840 m and cloud base height of 66 m. The main deviation in cloud top height stems mainly from the first half of the day where thick clouds prevail and the ground-based lidar is known to not be able to penetrate clouds beyond an optical thickness of 2-3 \citep{protat_statistical_2010}. Until 04:00 a four km thick anvil cloud is observed at Niamey. The anvil is the outflow of a convective core that passes Niamey between 04:00 and 12:00 followed by low mixed-phase clouds throughout the day with the interruption of a narrow convective core between 15:30 and 16:00. Although the input to the model only represents a two-dimensional view from space of the top of the atmosphere, IceCloudNet is reproducing the 3D structure of the cloud with only slight deviations. However, it can be observed that at 13:00 and 17:00, IceCloudNet's cloud cover has a gap, the latter is probably caused by the cirrus cloud that masks the signal from the lower cloud layer. Panels G and H provide a view on the 3D cloud structure at 00:12:41 representing a latitudinal and longitudinal cross section, respectively. The red dot on the x-axes marks the location of Niamey. The thick anvil cloud visible in the time resolved plots at Niamey is visible at the location of the red dot. In panel G another narrow active deep convective cloud can be observed at $\sim$17.5°N and the wide convective system that moves over Niamey between 04:00-12:00 is visible in H between 3°E and 8°E. IceCloudNet predictions enable studying the development and evolution of the convective cores and anvils of the \ac{mcs} in a four dimensional space. A rendering of the 3D predictions for this case study is provided in the video supplement. 


\subsubsection{Mindelo - 2024-01-16}

In the second case study, we analyze the IceCloudNet predictions at the location of the Mindelo Cloudnet station on Cabo Verde on 2024-01-16. Since the satellite missions underlying DARDAR recently finished their service, producing 3D cloud profiles for this date emphasizes the value of this project to the scientific community by producing cloud profiles beyond the lifetime of those satellite missions.

\begin{figure}[h]
  \centerline{\includegraphics[width=33pc]{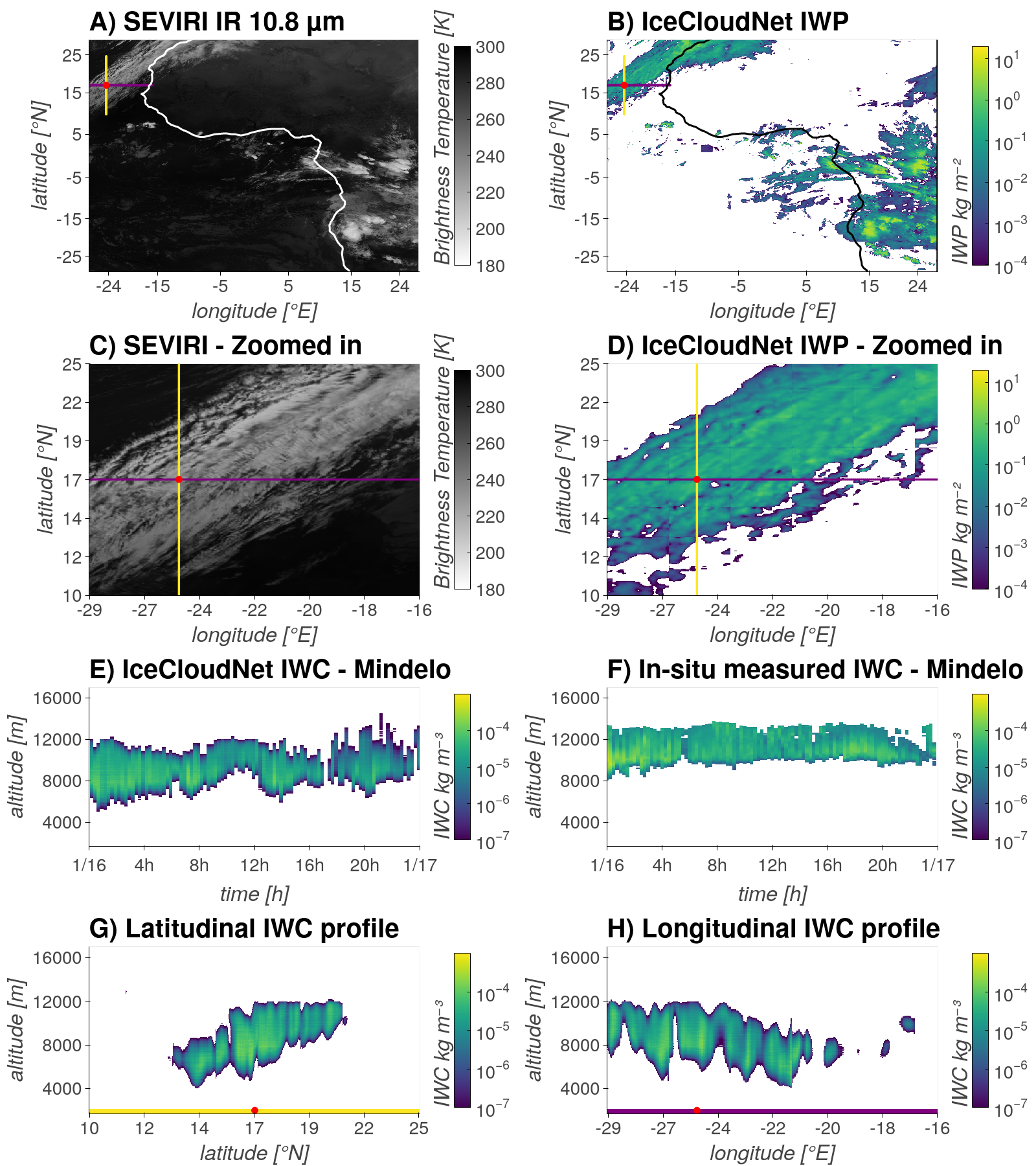}}
  \caption{As Fig. \ref{fig:case_study_niamey}, but for Cloudnet station in Mindelo on 2024-01-16.}
  \label{fig:case_study_mindelo}
\end{figure}
\FloatBarrier

Analogous to Fig. \ref{fig:case_study_niamey}, Fig. \ref{fig:case_study_mindelo} shows IceCloudNet predictions and SEVIRI input data for 2024-01-16. The scene shows an in-situ cirrus cloud with large horizontal extent. Again, IceCloudNet is able to represent cloud occurrence and extent throughout the day, with the deviations in cloud top height and base of 82m and 532m, respectively. The underestimation of \ac{iwc} and the smoothing effect towards the cloud edges by IceCloudNet can be observed in panel E.

\newpage

\section{Possible applications of IceCloudNet}\label{sec:use_cases}

In this section, a few exemplary use cases for the vertically and temporally resolved cloud ice reconstructions by IceCloudNet are outlined. IceCloudNet expands DARDAR observations in the spatial domain from cross sections (overpasses) to fully resolved three-dimensional cubes and from temporal snapshots to time-series data with a 15 minute resolution. It is thus now possible to take a perspective on the formation and temporal evolution of tropical mixed-phase and cirrus clouds.

\paragraph{Cloud ice origin} has been established as a way to classify cirrus clouds \citep{kramer_microphysics_2016} where in-situ cirrus form directly in the ice phase and liquid-origin cirrus are lifted to cirrus level from the mixed-phase regime. Using only Eulerian snapshots of active satellite instruments wrongly classifies anvil cirrus as in-situ \citep{gasparini_cirrus_2018}. \cite{wernli_trajectory-based_2016} applied an Lagrangian approach to study cirrus ice origin using backward trajectories of reanalysis data. With IceCloudNet it is now possible to move from reanalysis data to observational grade data and study cirrus ice origin in a Lagrangian setting. 

\paragraph{Aerosols} acting as \ac{inp} can influence the ice crystal formation mechanism and hence the cloud microphysical properties and radiative properties of cirrus and mixed-phase clouds. It remains a major subject of ongoing studies \citep{kanji_overview_2017} which aerosols can act as \ac{inp}, in which region, and which clouds are more sensitive to INP availability. \cite{froyd_dominant_2022} analysed the INP impact on cirrus properties where the authors combined trajectories of aerosols with microphysical modeling of cirrus clouds. A similar approach could be applied to the IceCloudNet dataset by co-locating aerosol trajectories and analyzing the predicted cloud ice properties of IceCloudNet's \ac{iwc} and \ac{nice} along trajectories. 

\paragraph{Meteorological properties} such as temperature and updraft velocities are mainly determining the cloud ice properties. \cite{jeggle_understanding_2023} combined backward trajectories of reanalysis data with DARDAR retrievals to study the impact of meteorological and aerosol drivers on cirrus microphysical properties. The main drawback of this approach is that it is not known in which stage of its development any given observed cloud is in, since DARDAR provides only an Eulerian perspective. This approach can now be extended by replacing DARDAR with IceCloudNet predictions allowing to follow the cloud properties throughout the trajectories and thus enabling a better view on cloud formation.

\paragraph{\ac{mcs} tracking} in geostationary satellite observation such as SEVIRI has been proven to be useful to improve understanding of storm tracking and the development of convection that precedes anvil cirrus in the tropics \citep{jones_lagrangian_2023}. A potential way forward is to co-locate the detected storm tracks with IceCloudNet data to analyze the microphysical properties from the development of the cumulus stage to the mature stage, and finally to the development of anvil cirrus in the decaying stage \citep{lohmann_introduction_2016}. 

\paragraph{High-resolution climate models} are able to resolve cloud-scale processes and promise improved climate and weather simulations ~\citep{satoh_global_2019}. To evaluate how well the simulations can represent cloud processes, a validation against observed cloud properties is required. IceCloudNet predictions provide a first step in this direction, offering a full vertical and time-resolved perspective for a large domain including the tropical Atlantic and Africa.

\paragraph{The recently launched EarthCare} satellite carrying multiple active and passive instruments succeeds CALIPSO and CloudSat as data source for vertical cloud profiles \citep{illingworth_earthcare_2015}. The new instruments will provide improved retrievals of cloud properties offering better observational constraints. A logical extension of IceCloudNet is to retrain the model on co-located EarthCare data and exploit the advantages of EarthCARE over its predecessors.

\section{Limitations and Potential for Improvements}\label{sec:limitations}

IceCloudNet predictions achieve strong agreement with independent test data both in predicting cloud occurrence and coverage (section~\ref{sec:results}.\ref{sec:results_cloud_occurrence}) and in predicting the magnitude of \ac{iwc} and \ac{nice}, especially for altitudes between 6 km and 15 km (section~\ref{sec:results}.\ref{sec:results_cloud_ice}). Also the exemplary comparison with ground-based stations supports the quality of IceCloudNet's prediction. Yet, we want to emphasize that IceCloudNet's predictions should be treated as synthetic data and some limitations taken into account when using the data to study cloud development. 

\subsection{Limitations}

One limitation of IceCloudNet is the data quality on which it was trained. As DARDAR is a retrieval product and hence does not actually measure cloud ice properties, resulting in inherent uncertainties, which can also transfer to IceCloudNet predictions. 
The insensitivity of SEVIRI channels to thin cirrus clouds that occur at high altitudes leads to a deteriorated performance of IceCloudNet at altitudes above 15 km, with a pronounced effect of \ac{nice}, hence we advise treating predictions at these altitudes with caution. Other data-induced limitations are the changing resolutions of SEVIRI depending on location and the temporal co-location bias described in section~\ref{sec:results}.\ref{sec:data_train}. 

IceCloudNet is a purely data-driven approach, which means that the results are not necessarily physically consistent. Without a discriminator module, IceCloudNet predicted \ac{iwc} $<$ 10$^{-6}$ kg m$^{-3}$ at all cloud edges independent of altitude. As \ac{iwc} is strongly dependent on temperature, these values do not occur in reality for clouds at lower altitudes. By introducing the discriminator to IceCloudNet, this blurring effect was mitigated, and the predictions in general are more similar to realistic cloud profiles now, but non-physical artefacts may still be produced in some cases.

\subsection{Future Development}

Including meteorological information such as temperature and vertical velocity could help IceCloudNet to learn and represent more consistent physical relationships in the training data. The current model could be additionally conditioned on the reanalysis data of meteorological variables, which could help the network to model important physical properties, such as the melting level and the homogeneous freezing level. Although modern reanalysis data, such as ERA5, have their own limitations, it could be worthwhile to evaluate their impact on IceCloudNet predictions. We also assume that passing several time steps of SEVIRI data as input prior to a DARDAR observation could improve the performance, as the network could benefit from the wider context of cloud formation and development in the input data. 

While we are confident in the design choices of the IceCloudNet architecture, we have not exhausted the space of commonly employed model architectures. Given the benefits offered by including the adversarial loss in IceCloudNet, an interesting approach would be the use of other powerful (but costly) generative models (such as the currently popular diffusion models~\citep{ho_denoising_2020}) to predict cloud profiles of higher fidelity. With new methods and techniques for training deep neural networks being constantly developed, we are certain that further performance improvements can be achieved with further improvements applied to the model architecture. Another major upgrade to IceCloudNet would be a quantification of the uncertainty of its predictions. 


\section{Conclusions}\label{sec:conclusions}

We introduce IceCloudNet, a new way to obtain high-quality vertically resolved predictions for microphysical properties of clouds containing ice with high spatio-temporal resolution and coverage. Trained on geostationary SEVIRI data and retrievals of actively sensed DARDAR data, our \ac{ml}-based approach achieves high agreement with independent test data and when evaluated against ground-based observations for exemplary case studies. The open-access dataset produced by IceCloudNet encompasses 10 years of vertically resolved \ac{iwc} and \ac{nice} data of clouds containing ice with a 3 km$\times$3 km$\times$240 m$\times$15 minute resolution in a spatial domain of 30°W to 30°E and 30°S to 30°N. The dataset increases the availability of vertical cloud profiles in years where DARDAR is available by over six orders of magnitude and provides vertical cloud profiles beyond the lifetime of recently-ended satellite missions underlying DARDAR. Supplying the scientific community with this new observational constraint will enable novel research on cloud ice formation and evolution from the single cloud scale to mesoscale cloud dynamics of \ac{mcs}. It will furthermore help to improve the understanding of the cloud microphysical processes and its drivers in cirrus and mixed-phase clouds by tracking and studying cloud properties through time and space. 

\clearpage
\acknowledgments
This research was supported by grants from the European Union’s Horizon 2020 research and innovation program iMIRACLI under Marie Sklodowska-Curie grant agreement No 860100 and Swiss National Supercomputing Centre (Centro Svizzero di Calcolo Scientifico, CSCS; project ID s1144). KJ is grateful for the opportunity to being hosted as a visiting researcher at ESA $\Phi$-lab during spring 2023 which led to the initiation of this research project. The authors thank Peter Naylor for his ideas on improving the model architecture, Tom Beucler for his comments on the manuscript and Sylvaine Ferrachat for her support on the dataset hosting. We thank the DKRZ, the WDCC, and Eileen Hertwig for the hosting of the IceCloudNet dataset. We are thankful for the availability of open-source Python software packages and source code, which greatly facilitated the development and analysis in this research. The authors declare that they have no conflict of interest.


%
%
\datastatement
10 years of IceCloudNet predictions from Meteosat SEVIRI images can be accessed from the World Data Center for Climate (\url{https://www.wdc-climate.de/ui/entry?acronym=IceCloudNet_3Drecon}). \ifarxiv Important note: with this preprint only a subset of the IceCloudNet data is available. The full 10 years of data will be available upon journal publication. \\
The co-located patches of SEVIRI and DARDAR data used for training IceCloudNet are not provided due to the licensing of SEVIRI. The original SEVIRI data can be accessed via the EUMETSAT Data Access Client Python library. The DARDAR-Nice data is available at \url{https://doi.org/10.25326/09}. \\
A repository containing the routines for creating the training data, training and evaluating the \ac{ml} model, pre-trained model weights, as well as a script for producing IceCloudNet data from SEVIRI input can be accessed at \url{https://github.com/tabularaza27/ice_cloud_net}. 


%






%



\bibliographystyle{ametsocV6}
\bibliography{references}

\end{document}